%% file: stripes4.tex
\documentclass[aps,prl,showpacs,twocolumn,floatfix,letterpaper,citeautoscript,10pt]{revtex4-1}

\usepackage[intlimits,sumlimits]{amsmath}
\usepackage{amsfonts,amssymb}
\usepackage{bm}
\usepackage{graphicx}
\usepackage{hyperref}
\usepackage[justification=centering]{caption}
\usepackage[vcentermath]{youngtab}
\usepackage{array}
\usepackage{multirow}

\newcommand{\be}{\begin{equation}}
\newcommand{\ee}{\end{equation}}

\hypersetup{%
pdftitle={Theory of Native Orientational Pinning in Quantum Hall Nematics},%
pdfauthor={Inti Sodemann, Allan H. MacDonald},%
pdfpagemode={UseNone},%
pdfstartview={FitH},%
breaklinks=true}

\begin{document}
\title{Theory of Orientational Pinning in Quantum Hall Nematics}

\author{I. Sodemann}
\affiliation{Department of Physics, University of Texas at Austin, Austin, Texas 78712}
\author{A. H. MacDonald}
\affiliation{Department of Physics, University of Texas at Austin, Austin, Texas 78712}

\date{\today}

\begin{abstract}
The orientation of the electron-nematic states discovered
in the quantum Hall regime of GaAs $[001]$ growth-direction
quantum wells is pinned by a weak native source of anisotropy.  
In this Letter we explain that this property, 
which has remained mysterious over more than a
decade of research, follows from the presence of both Rashba and Dresselhaus spin-orbit interactions.
The hard transport direction of the nematic state 
is determined by the relative sign of the Rashba and Dresselhaus coefficients,
and coincides with either the  $[110]$ or the $[1\bar{1}0]$ crystallographic direction. 
Our theoretical estimate of the pinning energy is in agreement with 
experimental studies of the competition between
native pinning and intentional pinning by an in-plane magnetic field.  
\end{abstract}
\maketitle

%%%%%%%%%%%%%%%%%%%%%%%%%%%%%%%%%%%%%%%%%%%%%%%%%%%%%%%%%%%%%%%%%%

{\em Introduction}---In the fractional quantum Hall regime two dimensional electron gases (2DEGs) host an unprecedented 
variety of strongly-correlated-electron states.  It is now established that, starting with the $n=2$ Landau level,
Hall resistivities no longer exhibit the sequence of plateaus that are present in the vicinity of
half-integer filling for $n=0$, and that longitudinal resistivities become strongly anisotropic~\cite{Lilly1999,Lilly1999a}
at low temperature instead of vanishing.  
A considerable body of experimental~\cite{Lilly1999,Lilly1999a,Du1999,*Pan1999,*Pan2000,*Willett2001a,*Cooper2001,*Cooper2002,*Zhu2002,*Cooper2004,*Sambandamurthy2008,*Zhu2009,*Kukushkin2011} and theoretical~\cite{Koulakov1996,*Moessner1996,*Fogler1996,*Fogler1997,*Fradkin1999,*Rezayi1999,*MacDonald2000,*Haldane2000,*Shibata2001,*Lopatnikova2001,Jungwirth1999,*Stanescu2000} evidence supports the view that these transport anisotropies 
signal the formation of electron nematics.  Locally electron nematics 
are unidirectional charge-density-wave (stripe) states, but because of disorder and thermal 
fluctuations they may lack long-range positional order.
Transport in nematics is {\em easy} along the stripe edges, which host chiral 
one-dimensional electron gases, but {\em hard} in the perpendicular direction.  
Experiment has shown that the easy transport direction is normally along the 
$[110]$ axis of GaAs quantum wells grown in the $[001]$ direction,
although a reorientation to the $[1\bar{1}0]$ axis has been observed when the density is increased above a critical value~\cite{Zhu2002}. In wide quantum wells  the stripe orientation has been observed to be different for 
majority and minority spin Landau levels~\cite{Liu2013}.  
The mechanism which selects these 
pinning directions is still undetermined after more than a decade of research.  

%%%%% FIG 1 %%%%%
\begin{figure}[t]
\includegraphics[width=2.75in]{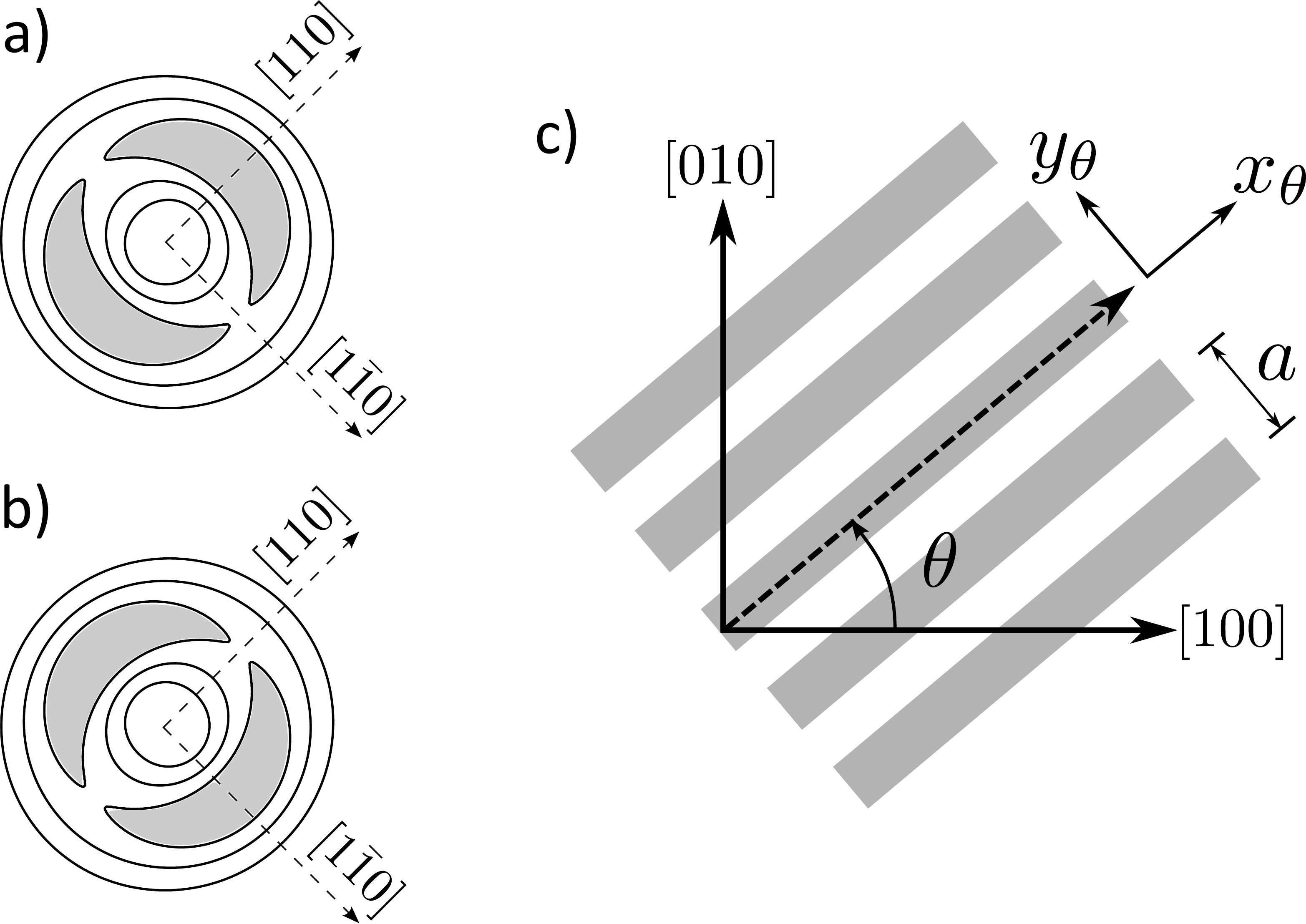}
\caption{Iso-probability contours of spin-orbit modified cyclotron wave-functions
in the $n=2$ Landau level: (a) $\epsilon_R=\epsilon_D=0.3$; (b) $\epsilon_R=-\epsilon_D=0.3$. 
The shaded area is the highest probability density region. 
c) Stripe state schematic showing areas occupied by orbit centers.
%The period, $a$, and the orientation, $\theta$, are are variational for the state.
}
\label{fig1}
\end{figure}
%%%%%%%%%%%%%%%%%

In this Letter we propose a theory in which orientational pinning follows from the combined presence of Rashba and Dresselhaus spin-orbit interactions~\footnote{We note that spin-orbit coupling has been argued to be responsible for the unconventional sequence of anisotropic states in two-dimensional hole systems of GaAs~\cite{Manfra2007}.}. 
Our theory posesses the following three desirable features: 
(i) the natural pinning axes are predicted to be either $[110]$ or $[1\bar{1}0]$ without parameter fine tunning; 
(ii) the pinning energy scale estimated from typical values of the Rashba and Dresselhaus interactions
in GaAs quantum wells is on the order of $\sim 10^{-7} eV$, in agreement with 
experiment~\cite{Lilly1999a,Jungwirth1999,Stanescu2000}; and (iii) it is falsifiable with a 
relatively simple experimental test. 
Previous proposed explanations for orientational pinning, including ones based on band mass anisotropy~\cite{Takhtamirov2000}, piezoelectricity~\cite{Fil2001,*Fil2002}, 
strain~\cite{Koduvayur2011}, and external potential modulations on lengths scales longer than the stripe period~\cite{Yoshioka2001,*Yang2003}, do not share these features.  
We predict that the relative sign of the Rashba and Dresselhaus constants determines 
which of the two natural axes, $[110]$ or $[1\bar{1}0]$, is chosen by the stripes. 
Stripes should therefore flip between $[110]$ and $[1\bar{1}0]$ orientations
when the sign of the perpendicular electric field which produces the Rashba effect is reversed.

{\em Landau-Levels and Spin-Orbit Interactions}---  
The starting point for our theory is an analysis of how spin-orbit interactions influence 
Landau level wavefunctions.  We specialize to the narrow quantum well limit in which the magnetic length $l \equiv(\hbar c/eB)^{1/2}$, greatly exceeds the well width $w$, which
affords several simplifications. In the narrow well limit we can restrict electronic states to the lowest subband, 
and neglect both finite-well-width corrections to the Coulomb interaction
and the cubic Dresselhaus term. 
The single-particle Hamiltonian is then $H=H_0+H_{\text{SO}}$ where $H_0$ contains the 
cyclotron and Zeeman energies,
\be\label{H0}
H_0=\frac{\pi ^2}{2 m^*}-\mu \cdot B=\hbar \omega _c(\hat{n}+1/2)+\frac{1}{2}|g^*| \mu _B B\sigma _z,
\ee
and $H_{SO}$ is the sum of the Rashba and Dresselhaus spin-orbit interaction terms~\cite{winkler2003spin}, 
\be\label{Hso}
H_{\text{SO}}=\alpha  \left(\sigma _x\pi _y-\sigma _y\pi _x\right)+\beta \left(\sigma _y\pi _y-\sigma _x\pi _x\right).
\ee
In Eq.~\eqref{H0} $\pi=p+eA/c$ is the mechanical momentum, $m^*$ and $g^*$ are the effective mass and $g$-factor of conduction electrons in GaAs, and $\omega_c=eB/m^*c$, where $B=\nabla \times A = - B \hat{z}$ is the 
magnetic field.  In Eq.~\eqref{Hso} $\alpha$ and $\beta$ are constants with units of velocity 
which specify the strengths of the Rashba and Dresselhaus interactions. 
It is convenient to define the dimensionless constants 
$\epsilon_R\equiv \sqrt{2} \alpha/l \omega _c$ and $\epsilon_D\equiv \sqrt{2} \beta / l \omega _c$, 
and to reexpress Eq.~\eqref{Hso} in terms of Landau level and 
spin raising and lowering operators, $a=l\left(\pi _x+i \pi _y\right)/\sqrt{2}\hbar$, $s_+=(\sigma_x+i \sigma_y)/2$ to 
obtain $H_{\text{SO}}= \hbar\omega _c \, (i \epsilon_R a^{\dagger }s_+-\epsilon_D a^{\dagger } s_- + h.c.) $.

Only the combined presence of Rashba and Dresselhauss spin-orbit interactions leads to anisotropy.
This property follows from the observation that the Rashba Hamiltonian commutes with sum of the orbital and spin angular momenta along the z-axis $L_z+\sigma_z/2$, whereas the Dresselhaus Hamiltonian commutes with their difference $L_z-\sigma_z/2$.
(Note that $L_z=b^\dagger b-a^\dagger a$, where $b$ is the guiding center
 lowering operator $b=\left(c_x+i c_y\right)/\sqrt{2}l$, and $c=r-l^2 \hat{z}\times \pi/\hbar$ is the 
classical cyclotron-orbit guiding center coordinate~\cite{cond-mat/9410047,giuliani2005quantum}.) 
When only one of the spin-orbit coupling terms is present, a spatial rotation 
compensated by the appropriate spin-rotation leaves the Hamiltonian invariant.  
This conclusion survives interactions because Coulomb coupling is 
spin-independent and isotropic.

For quantitative calculations we must evaluate the Landau level wave functions 
explicitly.  Provided that the dimensionless spin-orbit coupling constants $\epsilon_{R,D}$ are small 
we can safely use standard Raleigh-Schr\"oedinger perturbation theory.
%\be\label{rd}
%\epsilon_R\equiv\frac{\sqrt{2} \alpha }{l \omega _c}, \ \epsilon_D\equiv \frac{\sqrt{2} \beta }{l \omega _c}.
%\ee
When the spin-orbit Hamiltonian acts on an unperturbed wavefunction, it always reverses spin and changes
the Landau level index by $\pm 1$.
Second order processes therefore always preserve spin. 
The coherent addition of wave functions that describe cyclotron 
orbits with different radii yields anisotropic wavefunctions, as illustrated in Figs.~\ref{fig1}(a)-(b). 
Writing the eigenstates in the absence of spin-orbit interactions as $|n,\uparrow\rangle_0,|n,\downarrow\rangle_0$, the 
first and second order corrections to the wavefunctions (in the limit of zero Zeeman energy) are, 

\begin{gather}
\label{firstorderup}
|n,\uparrow \rangle_{1}=-i \epsilon_R \sqrt{n} \, |n-1,\downarrow\rangle_{1}+\epsilon_D \sqrt{n+1} \, |n+1,\downarrow\rangle_0,\\
\label{firstorderdown}
|n,\downarrow \rangle_1=-i \epsilon_R \sqrt{n+1} \, |n+1,\uparrow\rangle_0-\epsilon_D \sqrt{n} \, |n-1,\uparrow\rangle_0,
\end{gather}

and
\begin{widetext} 
\begin{equation}
\label{secondorder}
|n,\uparrow\rangle _2= \frac{i \epsilon_R \epsilon_D \sqrt{n(n-1)}}{2}\, |n-2,\uparrow\rangle_0 
-\frac{i \epsilon_R \epsilon_D \sqrt{(n+2)(n+1)}}{2} \, |n+2,\uparrow \rangle_0
 -  \frac{\epsilon_R^2 n + \epsilon_D^2 (n+1)}{2} \, |n,\uparrow\rangle_0.
 \end{equation}
 \end{widetext} 
%\begin{multline}
%|n,1\rangle _2=\frac{i \epsilon_R \epsilon_D \sqrt{n(n-1)}}{2(1+\epsilon_Z)}|n-2,1\rangle_0 \cdots\\
%-\frac{i \epsilon_R \epsilon_D \sqrt{(n+2)(n+1)}}{2(1-\epsilon_Z)}|n+2,1\rangle_0,
%\end{multline}
%\begin{multline}\label{rd2-}
%|n,-1\rangle _2=\frac{i \epsilon_R \epsilon_D \sqrt{n(n-1)}}{2(1+\epsilon_Z)}|n-2,-1\rangle_0 \cdots\\
%-\frac{i \epsilon_R \epsilon_D \sqrt{(n+2)(n+1)}}{2(1-\epsilon_Z)}|n+2,-1\rangle_0.
%\end{multline}
%The corresponding expressions for the perturbed $\downarrow$-states follow from obvious substitutions.  

\noindent $|n,\downarrow\rangle_2$ is obtained from the expression above by replacing $\uparrow$ by $\downarrow$ and interchanging $\epsilon_R\leftrightarrow\epsilon_D$. The finite Zeeman energy produces only weak corrections to the coefficients of Eqs.~\eqref{firstorderup},~\eqref{firstorderdown}, and~\eqref{secondorder}, proportional to its ratio to the cyclotron energy, 
which is small in GaAs except at extreme field tilt angles.  

{\em Form Factors and Anisotropy}---
We will assume that stripe states are always maximally spin-polarized.  With this simplification
the Hamiltonian is given up to a constant by the Coulomb interaction projected onto a
single spin sublevel $|n,\sigma\rangle$ that has been perturbed by spin-orbit interactions:
\be
\label{Projected} 
\bar{V}=\frac{1}{A}\sum_{i<j}\sum_{q\neq0} v_q |F_{n\sigma}(q)|^2 e^{iq(c_i-c_j)},
\ee
where $v_q = 2 \pi e^2/\epsilon q$ is the 2D Coulomb interaction and we have separated the 
2D position operator $r_i$ into guiding center $c_i$ and a cyclotron-orbit components using $r_i=c_i+l^2 \hat{z}\times\pi_i /\hbar$~\cite{cond-mat/9410047,giuliani2005quantum}.  Quantization of the cyclotron-motion replaces functions
of the mechanical momentum $\pi$ by expectation values that are responsible for the form factors 
\be\label{F}
F_{n\sigma}(q)\equiv\langle n, \sigma|e^{i  l^2 q \cdot (z\times\pi)/\hbar}| n, \sigma\rangle, 
\ee
in Eq.~\eqref{Projected}.  
The influence of spin-orbit coupling  appears in the modified form factor.
In the absence of spin-orbit coupling $F^0_n(q)=L_n( l^2|q|^2/2) \exp(- l^2|q|^2/4)$. 
Because electron-electron interactions are diagonal in spin, the leading corrections appear at  
second order.  From Eqs.~\eqref{firstorderup},~\eqref{firstorderdown}, and~\eqref{secondorder} we find that,
%\begin{equation}\label{F0}
%\langle n' |e^{iq\cdot z\times\pi}|n \rangle =
%\sqrt{\frac{n!}{n'!}}e^{-\frac{|q|^2}{4}} L_n^{n'-n}\left(\frac{|q|^2}{2}\right)\left(-\frac{q_x+iq_y}{\sqrt{2}}\right)^{n'-n},
%\end{equation}
\begin{widetext}
\begin{eqnarray} 
\label{formfactor}
F_{n\uparrow}^{}(q) &=& F_{n}^0(q)+n \epsilon_R^2 \left[F_{n-1}^0(q)-F_n^0(q)\right]+
(n+1) \epsilon_D^2 \left[F_{n+1}^0(q)-F_n^0(q)\right] \nonumber \\
&+& \frac{1}{2}\epsilon_R \epsilon_D \, q^2 l^2 \sin(2\theta_q) \, e^{-|q|^2l^2/4} 
 \left[2 L_{n-1}^2\left(\left.|q|^2l^2\right/2\right)-L_n^2\left(\left.|q|^2l^2\right/2\right)-L_{n-2}^2\left(\left.|q|^2l^2\right/2\right)\right].
\end{eqnarray}
\end{widetext} 
where $\theta_q$ is a momentum-space orientation angle.  
The final term in Eq~\eqref{formfactor} is responsible for anisotropy. 
Form factor contributions that are dependent on $\theta_q$ appear at any order in perturbation theory
only if both Rashba and Dresselhaus spin-orbit interactions are present.   
The form factor for $\downarrow$ states can be obtained from Eq.~\eqref{formfactor}, by interchanging
the Rashba and Dresselhaus coefficients $\epsilon_R \leftrightarrow \epsilon_D$.     

{\em Orientational Pinning}--- 
We are now in a position to estimate the pinning energy of the stripes. 
To construct the stripe state, we consider the rotated guiding center operator,
\be
c_{y}^\theta=-\text{sin}\theta \, c_x+\text{cos}\theta \, c_y.
\ee
The eigenstates of $c_{y}^\theta$ are localized as a function of $y^\theta=-x \text{sin}\theta+y \text{cos}\theta$,
but extended along the line $x^\theta=x \text{cos}\theta+y \text{sin}\theta$. 
We construct a single-Slater-determinant electron-nematic
trial wave function by occupying eigenstates of $c_{y}^\theta$ with eigenvalue $k$ inside the region $K$ 
defined by the periodically repeated strips depicted in Fig.~\ref{fig1}(c). 
The many-body wavefunction of a $(n,\sigma)$-Landau-level  stripe state 
that has its hard transport direction along $\hat{y}^\theta$, can therefore be written as, 
\be
|\Psi_{n,\sigma}(\theta,a)\rangle =\prod _{k\in K} C_{n\sigma k}^{\theta\dagger } |\Psi^0_{n,\sigma}\rangle , 
\ee
\noindent where $|\Psi^0_{n,\sigma}\rangle$ is a vacuum in which the lower Landau levels are 
are completely filled and $C_{n\sigma k}^{\theta\dagger}$ creates electrons $(n,\sigma)$ with $c_{y}^\theta$ 
eigenvalue $k$.  For a given valence Landau level filling factor $\nu$, 
the stripe state, $|\Psi_{n,\sigma}(\theta,a)\rangle$ has two variational parameters: $\theta$ which
characterizes the direction measured from the $[100]$ axis along which the stripes run, and the 
stripe period $a$. These two free parameters must be optimized to minimize the stripe energy. 

Except for the Landau levels being perturbed by spin orbit interactions, these variational wave functions are identical to those conventionally employed to perform Hartree-Fock studies of stripe states,  
in particular those which address the influence of an the in-plane magnetic field~\cite{Jungwirth1999}.
For given values of $\nu$, $a$, and $\theta$ the variational energies can be expressed as: 
\be
\label{energy} 
E=\frac{N_{\phi }}{4 \pi l^2}\sum_{n=-\infty }^{\infty } v_{HF}\left(\frac{2 n \pi }{a}\right)\left[\frac{\text{sin}(n \pi  \nu )}{n \pi }\right]^2,
\ee 
where $N_\phi=A/2\pi l^2$ is the orbital Landau level degeneracy and 
$v_{HF}(p)$ is the sum of a Hartree ($v_{H}(p)$) and a Fock ($v_{F}(p)$) contribution: 
\begin{eqnarray}
v_H(p) &=& v_q\left|F_{n,\sigma}(q)\right|{}^2_{q^\theta_x=0,q^\theta_y=p} , \nonumber \\
v_F(p) &=&-\frac{A}{N_{\phi }}\int \frac{d^2q}{(2 \pi )^2}v_q\left|F_{n,\sigma}(q)\right|{}^2e^{ip q^\theta_x}.
\end{eqnarray}
Here $(q^\theta_x,q^\theta_y)$ are measured along the stripe axes $(x^\theta,y^\theta)$ (see Fig.~\ref{fig1}(c)), and the Hartree potential is understood to vanish for $p=0$ (i.e. $v_H(p=0)=0$), to account for the nautralizing background charge.

The leading order anisotropic contribution to the energy in Eq.~\eqref{energy},
which is entirely due to spin-orbit coupling and 
comes from the anisotropic/isotropic cross term 
in $\left|F_{n,\sigma}(q)\right|{}^2$, has the form
\be\label{Eani}
E^{\text{ani}}(\theta,a,\nu)=N_\phi \ \text{sin}2\theta \ \varepsilon(a,\nu) \ \epsilon_R \epsilon_D \frac{e^2}{\epsilon l},
\ee
where $\varepsilon(a,\nu)$ is a dimensionless number.
Eq.~\eqref{Eani} is the key result of this study and 
predicts that the minimum of energy is reached when the stripes are aligned along the $[110]$ axis when 
$\varepsilon(a,\nu) \epsilon_R \epsilon_D < 0$ and along the $[1\bar{1}0]$ axis when 
$\varepsilon(a,\nu) \epsilon_R \epsilon_D > 0$.  
Numerical values for $\varepsilon$ are listed in Table~\ref{pinningtab}.

We can estimate spin-orbit parameters for the samples of Ref~\cite{Lilly1999a} from their 
carrier densities $n_0$ and well widths $w$ using a simple capacitor model: 
$\hbar \alpha=r^{6c6c}_{41}E^{\text{eff}}_z\sim r^{6c6c}_{41}4\pi e n_0/\epsilon$, and $\hbar\beta=b^{6c6c}_{41}\langle k_z^2\rangle\sim b^{6c6c}_{41}/w^2 \sim b^{6c6c}_{41}\left(4 \pi  n_0/a^*\right)^{2/3}$. Here $r^{6c6c}_{41}\approx 5.2e$\AA$^2$ and $b^{6c6c}_{41}\approx 27.6 e$V\AA$^3$ are material parameters for GaAs~\cite{winkler2003spin}, and $a^*\approx 103 {\rm \AA}$ is the effective Bohr radius of GaAs.   Using these estimates 
the stripe pinning energy scale is found to be $\epsilon_R \epsilon_D e^2/(\epsilon l) \sim 5.2\times10^{-7}eV\sim 6.1mK$ 
at $B \sim 1$T. 
This value agrees with that determined from experiments in which the preferred stripe orientation 
is changed by tilting the applied magnetic field away from the normal to the 2DEG plane~\cite{Lilly1999a,Jungwirth1999}.
Agreement with experiment for both qualitative and quantitative characteristics of the 
native pinning effect strongly suggests that we have identified the mechanism that is responsible. 

{\em Discussion}---
Our theory does not account for Landau-level mixing even though 
stripe states are normally studied experimentally at relatively weak magnetic fields.
We can partially assess its importance by comparing estimates made with 
bare Coulomb interactions with those in which inter-Landau level contributions to 
polarization functions are used to construct a  statically screened RPA Coulomb interaction. 
In this approximation, we replace the Coulomb interaction $v_q$ by
$ v^{RPA}_q= v_q/(1-v_q\chi_q)$ where 
where $\chi_q$~\cite{giuliani2005quantum},
\be\label{chi}
\chi_q=\frac{1}{2\pi l^2}\sum_{\sigma,\sigma',n' \ne n}\frac{f_{n\sigma}-f_{n'\sigma'}}{E_{n\sigma}-E_{n'\sigma'}}|F_{n\sigma,n'\sigma'}(q)|^2,
\ee
Here $f_{n\sigma}$, is a Fermi occupation factor which equals $1$ for the fully occupied Landau sublevels and $0$ for the fully empty ones, and it equals the fractional part of the total filling factor $\nu$ for the partially occupied Landau sublevel in which the stripe is constructed.  The form factors $F_{n\sigma,n'\sigma'}(q)$ in Eq.~\eqref{chi} are the off-diagonal generalizations of the density form factors defined in Eq.~\eqref{F}. 
This susceptibility applies for a translationally invariant state and therefore neglects modifications to 
screening arising from the stripe state itself. In practice one does not need to include the spin-orbit modifications to the energies or to the density form factors appearing in the density-density response function of Eq.~\eqref{chi} because they make a small relative contribution.
We see in Table~\ref{pinningtab} that at typical fields, screening reduces the 
estimated anisotropy energy by $\sim 10-20 \%$.  As long as other Landau level mixing 
effects, which cannot be accounted for simply by changing the effective interaction~\cite{Sodemann2013,*Peterson2013,*Simon2013}, 
have similar importance, our main conclusions should be reliable.   

\begin{table}
\caption{Stripe period $a_0$ and anisotropy energy parameter $\varepsilon$ (Eq.~\eqref{Eani}) 
for half-filled higher Landau levels. The calculations were carried out neglecting and including 
RPA screening evaluated in GaAs at $B\approx1T$.} 
\centering 
\begin{tabular}{c | c  c  c  c  c  c} 
\hline
\hline 
$\nu_{\text{total}}$ & \ $9/2$ & $11/2$ & $13/2$ & $15/2$ & $17/2$ & $19/2$ \\ [0.5ex] 
 $a_0/l$ & \ $6.19$ & $6.19$ & $7.20$ & $7.20$ & $8.08$ & $8.08$ \\ [0.5ex]
$\varepsilon$ & \ $-0.093$ & $-0.093$ & $-0.102$ & $-0.102$ & $-0.113$ & 
$-0.113$\\ [0.5ex]  
 $a^{\text{RPA}}_0/l$ & \ $6.51$ & $6.57$ & $7.51$ & $7.54$ & $8.37$ & $8.41$ \\ [0.5ex]
$\varepsilon^{\text{RPA}}$ & \ $-0.072$ & $-0.075$ & $-0.089$ & $-0.091$ & $-0.102$ & 
$-0.105$\\ [0.5ex]
\hline\hline 
\end{tabular}
\label{pinningtab} 
\end{table}

In Table~\ref{pinningtab} we compare stripe periods and anisotropy energies for 
several half-filled Landau levels.   We have found that $\varepsilon(a_0,\nu)$ is dominated 
by its exchange energy contribution and that it is negative with a typical value 
$\sim -0.1$ near the optimal stripe period $a_0$ at half-filled Landau levels. We have estimated the optimal stripe period $a_0$ in the 
absence of of spin-orbit coupling. The small modifications to stripe period arising from the spin-orbit coupling terms can be safely neglected because of the small values of $\epsilon_{R,D}$.  The main 
role of spin-orbit interactions is simply to choose the preferred stripe orientation.

%Landau level mixing makes the pinning parameter $\varepsilon(a_0)$ depend on the ratio of Coulomb to cyclotron energies, which we call $\kappa$,
%
%\be
%\kappa=\frac{e^2}{\epsilon l\hbar \omega_c}.
%\ee
%
%\noindent This dependence is however weak. In Table~\ref{pinningtab} we illustrate how this parameter only changes by about 10\% at $\kappa=2.5$, with respect to its value for the bare Coulomb interaction. This illustrates that the pinning mechanism is robust to details of the interelectron interaction.   

Since the pinning parameter, $\varepsilon(a_0,\nu)$, is negative, we predict that stripes 
pin along the $[110]$ axis when $\epsilon_R \epsilon_D > 0$, and
along the $[1\bar{1}0]$ axis when $\epsilon_R \epsilon_D < 0$.
Because $\epsilon_{R}$ is an odd function of the effective 
electric field associated with structural inversion asymmetry whereas $\epsilon_{D}$ 
is even, we also predict that $\text{sign}(\epsilon_R \epsilon_D)=\text{sign}(E^{\text{eff}}_z)$.
Some caution must be exercised in applying this last conclusion since the effective electric fields are non-trivial~\cite{winkler2003spin}. 
Experiments have revealed that the Rashba constant can be finite even in a 
nominally symmetric quantum wells~\cite{Koralek2009}. 
Therefore, a test of the conclusion that the stripes rotate upon a change of sign of this effective electric field
should ideally be accompanied by an independent determination of the sign of the Rashba and Dresselhaus spin-orbit coupling 
constants.

Our theory has been specialized to the case in which the quantum well width is much smaller than the magnetic length $w\ll l \approx 26nm/\sqrt{B[T]}$. The widths of the wells employed in early observations of the stripes~\cite{Lilly1999,Lilly1999a}, can be estimated to be on the order of $10nm$ using the simple capacitor model. Thus it is reasonable to expect that our theory is not accurate at a quantitative level for these experiments. One of the terms that is neglected in the narrow well limit, and that could have a significant impact on the stripe pinning is the cubic Dresselhaus term, which alone breaks rotational symmetry. Additionally we have not explored the interplay of spin-orbit coupling and in-plane magnetic fields which may be important 
in wider quantum wells. 

Our study underscores the importance of the rotational symmetry breaking induced by spin-orbit coupling.  At moderate magnetic fields ($B\sim 1T$), the Zeeman energy scale is typically of the same order of magnitude of the spin-orbit coupling terms $\sim 10^{-5} eV$ in GaAs. The interplay of these terms remains a relatively unexplored subject, and should be of special important for situations where broken rotational invariance plays a role.

\begin{acknowledgements}

We are grateful to A. Croxall, J. Eisenstein, M. Fogler, B. Frie\ss, Y. Liu and M. Shayegan for estimulating discussions. This work was supported by the DOE Division of Materials Sciences and Engineering under grant DE-FG03-02ER45958 and by the 
Welch foundation under grant TBF1473.

\end{acknowledgements}

%%%%%%%%%%%%%%%%%%%%%%% References
%%%%%%%%%%%%%%%%%%%%%%%%%%
\input{stripes4.bbl}
%\bibliography{stripes}

\end{document}

%% file: stripes4.bbl
%merlin.mbs apsrev4-1.bst 2010-07-25 4.21a (PWD, AO, DPC) hacked
%Control: key (0)
%Control: author (8) initials jnrlst
%Control: editor formatted (1) identically to author
%Control: production of article title (-1) disabled
%Control: page (0) single
%Control: year (1) truncated
%Control: production of eprint (0) enabled
%